\documentstyle[prd,aps,preprint,psfig]{revtex}
\draft
\newcommand{\BR}{{\cal B}}
\newcommand{\eff}{\varepsilon}

\newcommand{\psp}{\psi(2S)}
\newcommand{\jpsi}{J/\psi}
\newcommand{\etac}{\eta_c}
\newcommand{\chicJ}{\chi_{cJ}}
\newcommand{\chicz}{\chi_{c0}}
\newcommand{\chico}{\chi_{c1}}
\newcommand{\chict}{\chi_{c2}}

\newcommand{\psipp}{\pi^+\pi^- J/\psi}

\newcommand{\pip}{\pi^+}
\newcommand{\pim}{\pi^-}

\newcommand{\pp}{\pi^+\pi^-}
\newcommand{\kk}{K^+K^-}
\newcommand{\ppb}{p\bar{p}}

\newcommand{\pppp}{\pi^+\pi^-\pi^+\pi^-}
\newcommand{\ppkk}{\pi^+\pi^-K^+K^-}
\newcommand{\pppr}{\pi^+\pi^-p\bar{p}}
\newcommand{\kkkk}{K^+K^-K^+K^-}
\newcommand{\kskp}{K^0_s K^+ \pi^- + c.c.}

\newcommand{\ksks}{K^0_s K^0_s}
\newcommand{\phiphi}{\phi\phi}
\newcommand{\gpppp}{\gamma \pi^+\pi^-\pi^+\pi^-}
\newcommand{\gppkk}{\gamma \pi^+\pi^-K^+K^-}

\newcommand{\gkkkk}{\gamma K^+K^-K^+K^-}
\newcommand{\gkskp}{\gamma K^0_s K^+ \pi^- + c.c.}

\newcommand{\gksks}{\gamma K^0_s K^0_s}

\newcommand{\tpp}{3(\pi^+\pi^-)}

\newcommand{\ra}{\rightarrow}

\newcommand{\psito}{J/\psi \rightarrow }
\newcommand{\pspto}{\psi(2S) \rightarrow }
\newcommand{\chicJto}{\chi_{cJ} \rightarrow }
\newcommand{\chiczto}{\chi_{c0} \rightarrow }
\newcommand{\chicoto}{\chi_{c1} \rightarrow }
\newcommand{\chictto}{\chi_{c2} \rightarrow }
\newcommand{\bfg}{\begin{figure}}
\newcommand{\efg}{\end{figure}}
\newcommand{\bitm}{\begin{itemize}}
\newcommand{\eitm}{\end{itemize}}
\newcommand{\bnum}{\begin{enumerate}}
\newcommand{\enum}{\end{enumerate}}
\newcommand{\btbl}{\begin{table}}
\newcommand{\etbl}{\end{table}}
\newcommand{\btbu}{\begin{tabular}}
\newcommand{\etbu}{\end{tabular}}
\begin{document}

\preprint{\vbox{\hbox{BIHEP-EP1-98-05}}}

\title{ Study of the Hadronic Decays of $\chi_{c}$ States }
\author{
J.~Z.~Bai,$^1$   Y.~Ban,$^5$      J.~G.~Bian,$^1$
I.~Blum,$^{12}$ 
G.~P.~Chen,$^1$  H.~F.~Chen,$^{11}$  
J.~Chen,$^3$ 
J.~C.~Chen,$^1$  Y.~Chen,$^1$ Y.~B.~Chen,$^1$
% ...
\cite{atNU0}
Y.~Y.~Wang,$^1$  
M.~Weaver,$^2$
C.~L.~Wei,$^1$   Y.~G.~Wu,$^1$
D.~M.~Xi,$^1$    X.~M.~Xia,$^1$   P.~P.~Xie,$^1$    Y.~Xie,$^1$
Y.~H.~Xie,$^1$   G.~F.~Xu,$^1$    S.~T.~Xue,$^1$
J.~Yan,$^1$      W.~G.~Yan,$^1$   C.~M.~Yang,$^1$   C.~Y.~Yang,$^1$
J.~Yang,$^1$     
W.~Yang,$^3$
X.~F.~Yang,$^1$  M.~H.~Ye,$^1$    S.~W.~Ye,$^{11}$
Y.~X.~Ye,$^{11}$   C.~S.~Yu,$^1$    C.~X.~Yu,$^1$     G.~W.~Yu,$^1$
Y.~H.~Yu,$^4$    Z.~Q.~Yu,$^1$    C.~Z.~Yuan,$^{13,1}$   Y.~Yuan,$^1$
B.~Y.~Zhang,$^1$ C.~C.~Zhang,$^1$ D.~H.~Zhang,$^1$  Dehong~zhang,$^1$
H.~L.~Zhang,$^1$ J.~Zhang,$^1$    J.~W.~Zhang,$^1$  L.~S.~Zhang,$^1$
Q.~J.~Zhang,$^1$ S.~Q.~Zhang,$^1$ X.~Y.~Zhang,$^6$  Y.~Y.~Zhang,$^1$
D.~X.~Zhao,$^1$  H.~W.~Zhao,$^1$  Jiawei~Zhao,$^{11}$ J.~W.~Zhao,$^1$
M.~Zhao,$^1$     W.~R.~Zhao,$^1$  Z.~G.~Zhao,$^1$   J.~P.~Zheng,$^1$
L.~S.~Zheng,$^1$ Z.~P.~Zheng,$^1$ B.~Q.~Zhou,$^1$   G.~P.~Zhou,$^1$
H.~S.~Zhou,$^1$  L.~Zhou,$^1$     K.~J.~Zhu,$^1$    Q.~M.~Zhu,$^1$
Y.~C.~Zhu,$^1$   Y.~S.~Zhu,$^1$ and  B.~A.~Zhuang$^1$
\\ (BES Collaboration)}
\address{
$^1$Institute of High Energy Physics, Beijing 100039, People's Republic of
 China\\
$^2$California Institute of Technology, Pasadena, California 91125\\
$^3$Colorado State University, Fort Collins, Colorado 80523\\
$^4$Hangzhou University, Hangzhou 310028, People's Republic of China\\
$^5$Peking University, Beijing 100871, People's Republic of China\\
$^6$Shandong University, Jinan 250100, People's Republic of China\\
$^7$Shanghai Jiaotong University, Shanghai 200030, People's Republic of China\\
$^8$Stanford Linear Accelerator Center, Stanford, California 94309\\
$^9$University of Hawaii, Honolulu, Hawaii 96822\\
$^{10}$University of California at Irvine, Irvine, California 92717\\
$^{11}$University of Science and Technology of China, Hefei 230026,
People's Republic of China\\
$^{12}$University of Texas at Dallas, Richardson, Texas 75083-0688\\
$^{13}$China Center of Advanced Science and Technology (World Laboratory),
Beijing 100080, People's Republic of China}
\date{\today}
\maketitle

\begin{abstract}
Hadronic decays of the P-wave spin-triplet charmonium states
$\chicJ (J=0,1,2)$ are studied using a sample of $\psp$ decays
collected by the BES detector operating at the BEPC storage ring. 
Branching fractions for the decays $\chicoto \kskp$, $\chiczto \ksks$, 
$\chictto \ksks$, $\chiczto \phiphi$, $\chictto \phiphi$ 
and $\chicJ \ra \kkkk$ are measured for the first time, and
those for $\chicJto \pppp$, $\chicJto \ppkk$, 
$\chicJto \pppr$ and $\chicJto \tpp$ are measured with 
improved precision.  In addition, we determine the masses
of the $\chi_{c0}$ and $\eta_c$ to be 
$M_{\chi_{c0}}=3414.1 \pm 0.6(stat) \pm 0.8 (sys)$~MeV and
$M_{\eta_c}=2975.8 \pm 3.9(stat) \pm 1.2 (sys)$~MeV.
\end{abstract}


\section{Introduction}

The P-wave spin-triplet charmonium states were originally
observed~\cite{chic_obs} in radiative decays of the $\psp$
soon after the discovery of the $\jpsi$ and $\psp$ resonances.
A number of decay modes of these states have been observed
and branching fractions reported~\cite{pdg}.  Most of
the existing results are from the Mark~I experiment, 
which had a data
sample  of 0.33 million $\psp$ decays~\cite{tanenbaum}.
Because the photon capabilities of the Mark~I detector 
were limited, the detection of the
photon from the $\psp\rightarrow\gamma\chicJ$
process was not required, and 
one constraint kinematic fits were used to 
reconstruct the final states.

Recently there has been a renewed interest in the P-wave
charmonium states.  Since in lowest-order perturbative QCD
the $\chi_{c0}$ and $\chi_{c2}$ decay via the annihilation of
their constituent $c\bar{c}$ quarks into two gluons,
followed by the hadronization of the gluons into light
mesons and baryons, these decays are expected to be
similar to those of a bound $gg$ state; a detailed
knowledge of the hadronic decays of the $\chi_{c0}$
and $\chi_{c2}$  may provide an understanding of 
the decay patterns of glueball states that will help
in their identification.

The mass differences between the three $\chi_c$ states
provide information on the spin-orbit and tensor interactions
in non-relativistic potential models and lattice QCD calculations.
The masses of the $\chi_{c1}$ and $\chi_{c2}$ 
have been precisely determined (to a level of $\sim\pm 0.12$~MeV)
by Fermilab experiment E760~\cite{E760a} using
the line shape measured in the
$p\bar{p}\ra \chi_{c1,2}$ formation reaction.  In contrast,
the $\chi_{c0}$ mass is much more poorly known; the PDG average 
for $M_{\chicz}$ has an uncertainty of $\pm 2.8$~MeV~\cite{pdg}.

In this paper, we report the analyses of
all-charged-track final states from $\chicJ$ decays,
including $\pppp$, $\ppkk$, $\pppr$, $\kkkk$, $\kskp$
and $\tpp$. The results for 
$\chicJ$ decays into  $\pp$, $\kk$ and $\ppb$ have been reported
elsewhere~\cite{width}. 
We use the combined invariant mass distribution from all 
of the channels under study to determine the $\chi_{c0}$
mass with improved precision.  

A byproduct of this analysis is a determination of the mass 
of the $\etac$.  This is of interest because the 
$M_{J/\psi}-M_{\eta_c}$ mass difference measures the strength of
the hyperfine splitting term in heavy quark interactions.
However, in spite of a number of measurements,
the current experimental value of $M_{\eta_c}$ 
remains ambiguous:  
the PDG~\cite{pdg} average 
is based on a fit to seven measurements with poor internal  
consistency~\cite{E760,dm2,mk3} and the confidence level 
of the fit is only 0.001.  A recent
measurement from E760~\cite{E760}
disagrees with the value reported by the DM2
group~\cite{dm2}
by almost four standard deviations.  Additional
measurements may help clarify the situation. 

The data used for the analysis reported here were taken with the
BES detector at the BEPC storage ring
at a center-of-mass energy corresponding to $M_{\psp}$.
The data sample corresponds to a total of $(3.79 \pm 0.31)\times 10^6$
$\psp$ decays, as determined from the observed number of inclusive
$\pspto \psipp$ decays~\cite{npsp}.

\section{The BES Detector}

BES is a conventional solenoidal magnet detector that is
described in detail in Ref.~\cite{bes}. A four-layer central
drift chamber (CDC) surrounding the beampipe provides trigger
information. A forty-layer cylindrical main drift chamber (MDC), located
radially outside the CDC, provides trajectory and energy loss
($dE/dx$) information for charged tracks over $85\%$ of the
total solid angle.  The momentum resolution is
$\sigma _p/p = 0.017 \sqrt{1+p^2}$ ($p$ in $\hbox{\rm GeV}/c$),
and the $dE/dx$ resolution for hadron tracks is $\sim 11\%$.
An array of 48 scintillation counters surrounding the MDC  measures
the time-of-flight (TOF) of charged tracks with a resolution of
$\sim 450$ ps for hadrons.  Radially outside of the TOF system is a 12
radiation length thick, lead-gas barrel shower counter (BSC) operating 
in the limited streamer mode.  This device covers
$\sim 80\%$ of the total solid
angle and measures the energies of electrons and photons 
with an energy resolution of $\sigma_E/E=22\%/\sqrt{E}$ ($E$
in GeV).  Outside the BSC is a solenoid, which
provides a 0.4~Tesla magnetic field over the tracking volume.
An iron flux return is instrumented with three double layers 
of  counters that identify muons of momentum greater than 0.5~GeV/c.

\section{Monte Carlo}

We use Monte Carlo simulated events to 
determine the detection efficiency
($\eff$) and the mass resolution ($\sigma_{res}$)
for each channel analyzed.
The Monte Carlo program (MC) generates events of the type
$\psp\rightarrow\gamma\chi_{cJ}$ under the assumption
that these processes are pure $E1$ 
transitions~\cite{tanenbaum,e1}:
the photon polar angle distributions are
$1+\cos^2\theta~~(\chi_{c0})$, 
$1-\frac{1}{3}\cos^2\theta~~(\chi_{c1})$ and
$1+\frac{1}{13}\cos^2\theta~~(\chi_{c2})$.
Multihadronic $\chi_{cJ}$ decays are simulated using
phase space distributions.  For each channel, either
10000 or 5000 events are generated, 
depending on the numbers of events for the corresponding
mode that are observed in the data sample.

\section{Event selection}

\subsection{Photon Identification}

\begin{figure}[htbp]
\centerline{\hbox{
\psfig{file=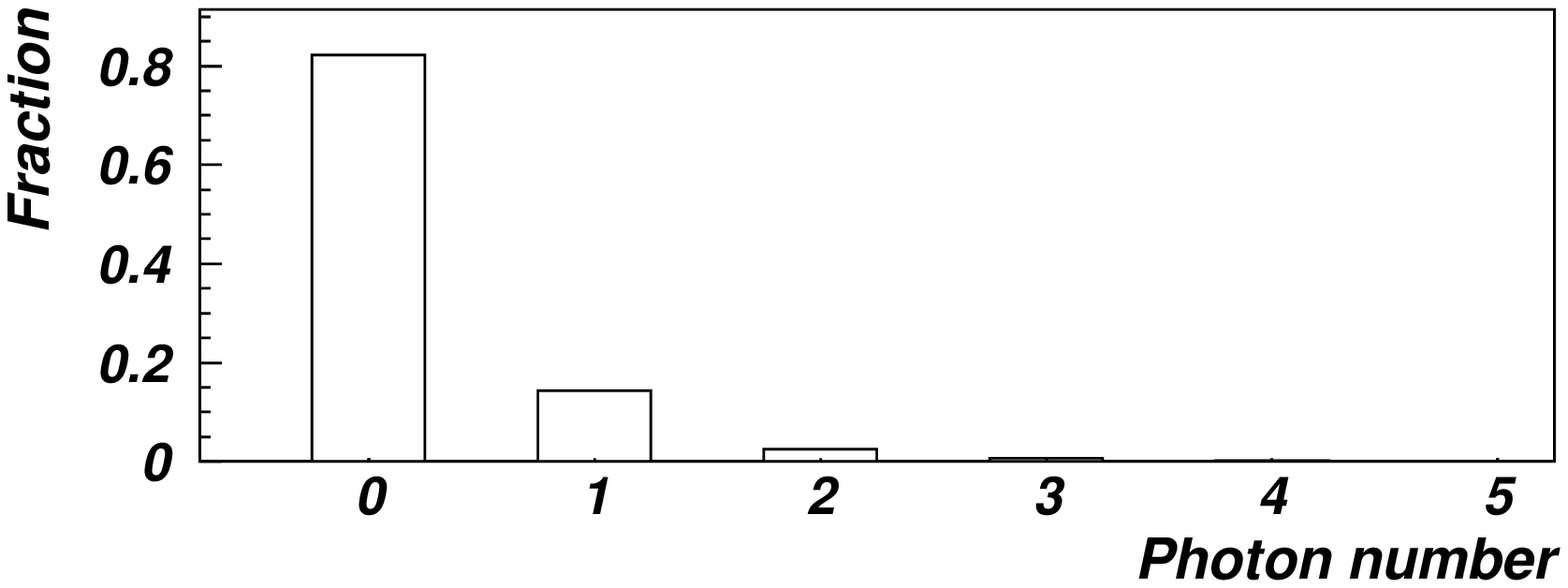,width=6.0cm}}}
\caption{The distribution of the
number of photon candidates found in 
$\pspto \pppp$ and $\pspto \ppkk$ events.}
\label{ngm}
\end{figure}

A neutral cluster is considered to be a photon candidate when the
angle in the $xy$ plane between the nearest
charged track and the cluster is greater than $15^{\circ}$, the first
hit is in the beginning 6 radiation lengths, and the difference between
the angle of the cluster development
direction in the BSC and the photon 
emission direction is less than $37^{\circ}$. 
When these selection criteria
are applied to kinematically selected samples of
$\psp \rightarrow \pi^{+}\pi^{-}\pi^{+}\pi^{-}$ and
$\psp \rightarrow \pi^{+}\pi^{-}K^{+}K^{-}$ events, fewer than
20\% of the events have $\gamma$ candidates, which
indicates that the fake-photon rejection ability
is adequate (see Fig.\ \ref{ngm}). 
The number of photon candidates in an event is limited to
four or less.  The photon candidate with the
largest energy deposit in the BSC is treated as the photon 
radiated from $\psp$ and used in a 
four-constraint kinematic fit to the hypothesis 
$\pspto \gamma + \hbox{charged tracks}$.

\subsection{Charged Particle Identification}

Each charged track is required to be well fit to a three-dimensional
helix and be in the polar angle region $|\cos\theta_{MDC}|<0.8$. 
For each track, the TOF and $dE/dx$ measurements are 
used to calculate $\chi^2$ values and the corresponding 
confidence levels to the hypotheses that the particle is a pion, 
kaon and proton ($Prob_{\pi},Prob_{K},Prob_{p}$).
The reliability of the confidence level assignments
is verified using a sample of $\psp \rightarrow \pi^+\pi^-J/\psi$, 
$J/\psi \rightarrow \rho \pi$ and $J/\psi \rightarrow \kk$ events, 
where the particle identification  confidence levels 
($ProbID$) of the tracks in different momentum ranges 
are found to be distributed uniformly between zero and one
as expected~\cite{jpsikk}.
Typically the $ProbID$ value of each track for a
given decay hypothesis is required to be 
greater than 1\% in our analysis.

\subsection{Event Selection Criteria}

For all decay channels, the candidate events are required
to satisfy the following selection criteria:
\begin{enumerate}
\item   The number of charged tracks is required to be four
        or six with net charge zero.
\item   The maximum number of neutral clusters in an event is eight, 
        and the number of photon candidates remaining
        after the application of the photon selection 
	is required to be four or less.
\item   The sum of the momenta of the lowest momentum $\pip$ and $\pim$
        tracks is required to be greater than 550~MeV; this removes
        contamination from $\psp \rightarrow \pi^+\pi^-J/\psi$ events.
\item   The $\chi^2$ probability for a four-constraint kinematic
        fit to the decay hypothesis is greater than 0.01.
\item   The particle identification assignment of each charged track is
        $ProbID > 0.01$.
\end{enumerate}

\subsubsection{$\gamma \pi^+\pi^-\pi^+\pi^-$ and
               $\gamma \pi^+\pi^-K^+K^-$}

A combined probability of the four-constraint kinematic fit and
particle identification information 
is used to separate $\gamma \pi^+\pi^-\pi^+\pi^-$ and 
the different particle assignments for the
 $\gamma \pi^+\pi^-K^+K^-$ final states.
This combined probability, $Prob_{all}$, is defined as
\[ Prob_{all}=Prob(\chi^{2}_{\hbox{all}}, \hbox{\em ndf}_{\hbox{all}}), \]
where
$\chi^{2}_{\hbox{all}}$ is the sum of the $\chi^2$ values from the
four-constraint kinematic fit and those from each of the four particle
identification assignments,
and
$\hbox{\em ndf}_{\hbox{all}}$
is the corresponding total number of degrees of the freedom used in the 
$\chi^2$ determinations.
The particle assignment with the largest $Prob_{all}$ is selected,
and further cuts on the kinematic fit probability and 
particle identification probability are imposed.

\begin{figure}[htbp]
\centerline{\hbox{
\psfig{file=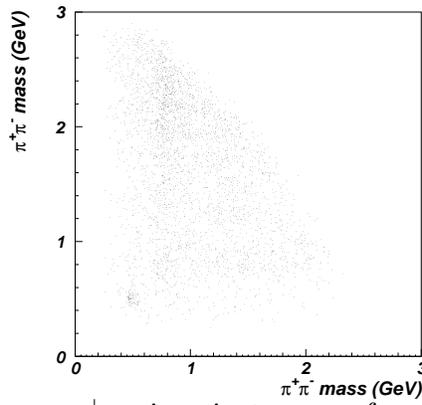,width=5.5cm}}}
\caption{A scatterplot of $\pip\pim$ vs $\pip\pim$ invariant
masses for selected $\gpppp$ events
         (two entries per event).}
\label{mks-mks}
\end{figure}

Figure\ \ref{mks-mks} shows a scatterplot
of $\pi^+\pi^-$ vs $\pi^+\pi^-$  invariant masses for
events with a $\pppp$ mass between 3.2 and 
3.6~GeV. The cluster of events in the lower left-hand corner
indicates the presence of a $\ksks$ signal.  A fit 
of a Gaussian function to the $\pip\pim$
mass distribution gives a peak mass at $499.3\pm 1.2$~MeV and
a width $\sigma = 11.8 \pm 1.0$~MeV that is
consistent with the MC expectation for the mass resolution. 
We select $\gamma K^0_s K^0_s$ candidates by requiring the mass of 
both $\pi^+\pi^-$ combinations in the event to be within $\pm2\sigma$ 
of the nominal $K^0_s$ mass. 

\begin{figure}[htbp]
\centerline{
\psfig{file=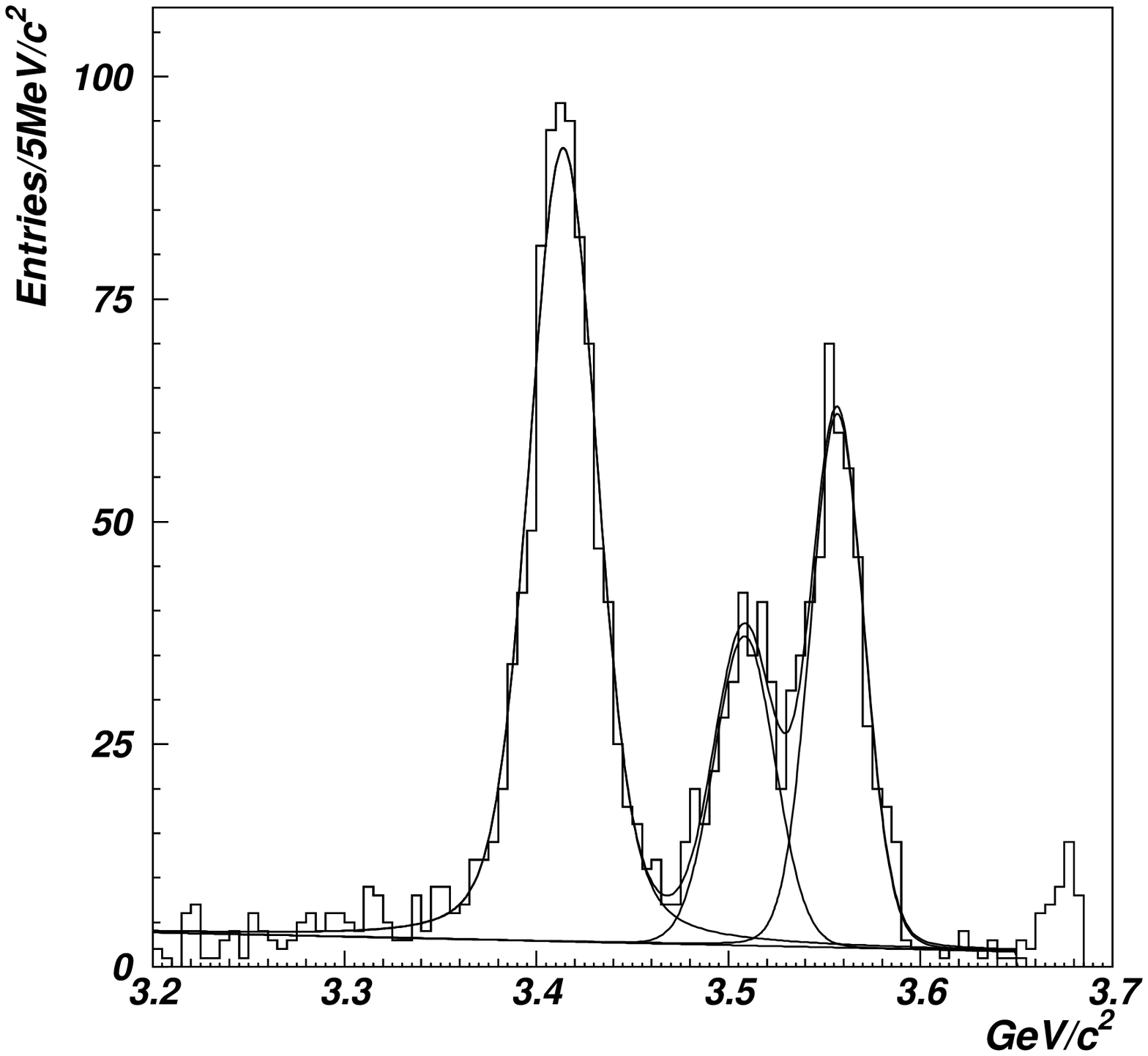,width=5.5cm}}
\caption{The $\pi^+\pi^-\pi^+\pi^-$ invariant mass distribution. 
The smooth curve is the result of a fit described in the text.}
\label{mallpppp}
\end{figure}

\begin{figure}[htbp]
\centerline{
\psfig{file=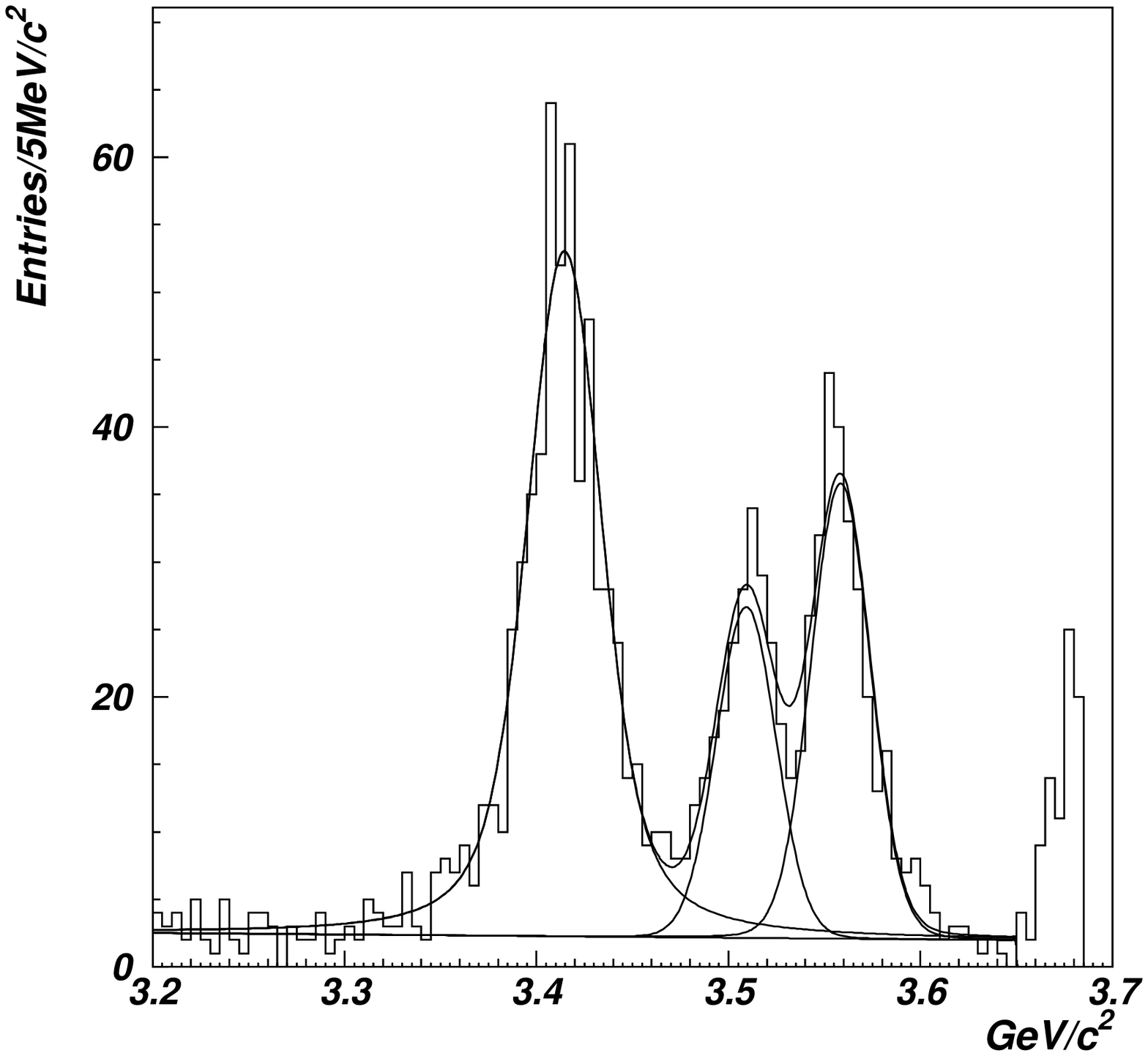,width=5.5cm}}
\caption{The $\ppkk$ invariant mass distribution.  The smooth curve
is the result of a fit described in the text.}
\label{mallppkk}
\end{figure}

\begin{figure}[htbp]
\centerline{
\psfig{file=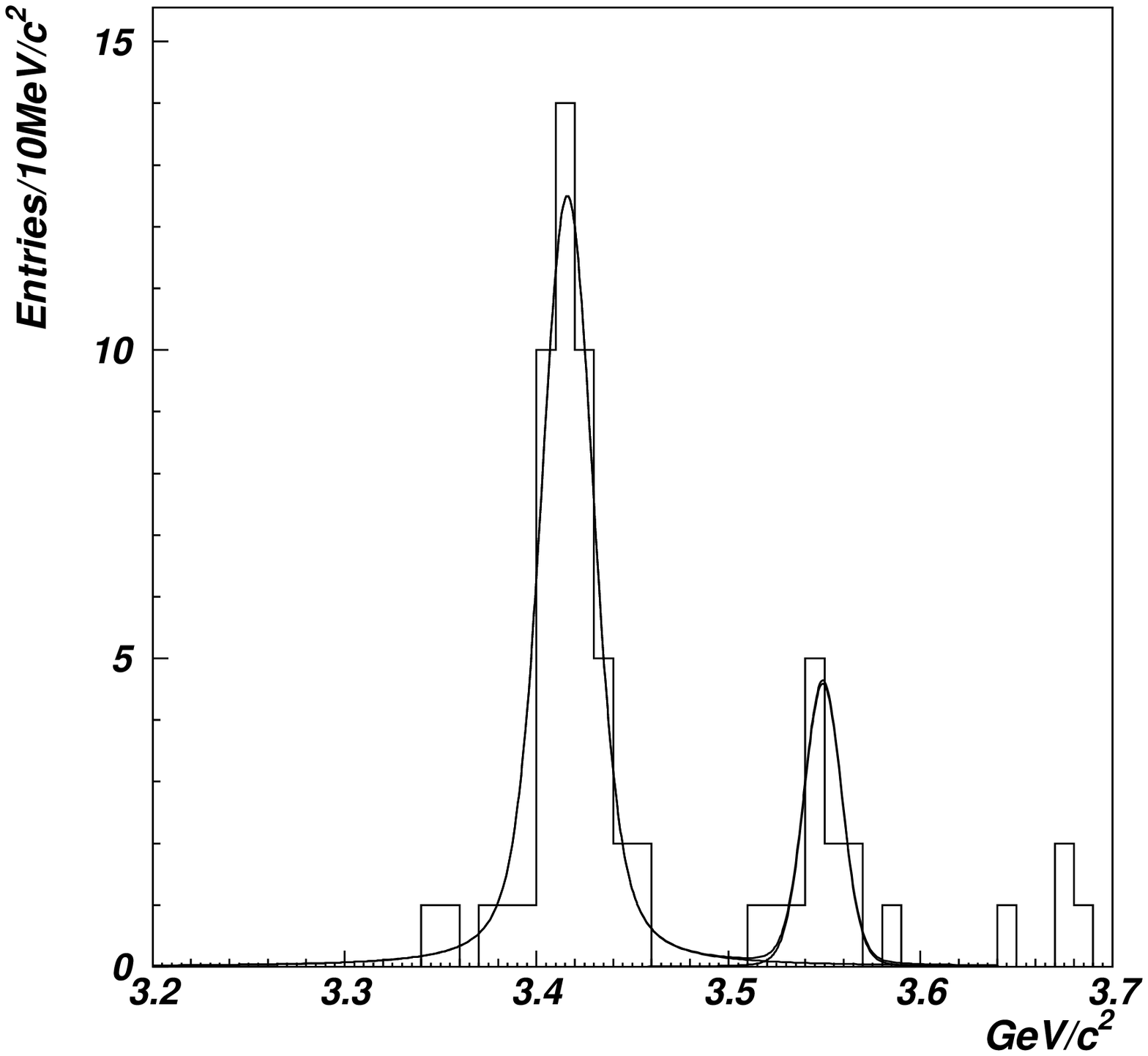,width=5.5cm}}
\caption{The $\ksks$ invariant mass distribution.
The smooth curve is the result of a fit described in the text.}
\label{mallksks}
\end{figure}

The invariant mass distributions for the $\pppp$, 
$\ppkk$ and $\ksks$ events that survive all the selection
requirements are shown in Figs.\ \ref{mallpppp}, 
\ref{mallppkk} and \ref{mallksks}.  There are peaks corresponding
to the $\chicJ$ states in each of the plots. (The high mass peaks
in Figs.\ \ref{mallpppp} and \ref{mallppkk} correspond to the
$\psp$ decays to all charged tracks final states that are
kinematically fit with a fake low-energy photon.)

We fit the $\pppp$, $\ppkk$ or $\ksks$ invariant mass distribution
between 3.20 and 3.65~GeV  with three Breit-Wigner resonances
convoluted with Gaussian mass resolution functions and
a linear background shape using an unbinned maximum likelihood method. 
In the fit, the mass resolutions are fixed to their MC-determined values 
and the widths of the $\chi_{c1}$ and $\chi_{c2}$ are fixed to the
PDG average values of of 0.88 and 2.00~MeV~\cite{pdg}, respectively.
The results of the fit 
are listed in Table\ \ref{tab-ppkk-res}
and shown in Figs.\ \ref{mallpppp}, \ref{mallppkk} 
and \ref{mallksks}. Table\ \ref{tab-ppkk-res} also lists
the MC-determined efficiencies and mass resolutions.

\subsubsection{$\gamma \pi^+\pi^-p \bar{p}$}

If one of the four tracks is identified as a proton or antiproton,
the event is assumed to be $\gamma \pi^+\pi^-p \bar{p}$.  We
assign probabilities to
the remaining particle assignment using the same technique
that was used for $\ppkk$ decays;
the combination with the highest probability is selected.

\begin{figure}[htbp]
\centerline{
\psfig{file=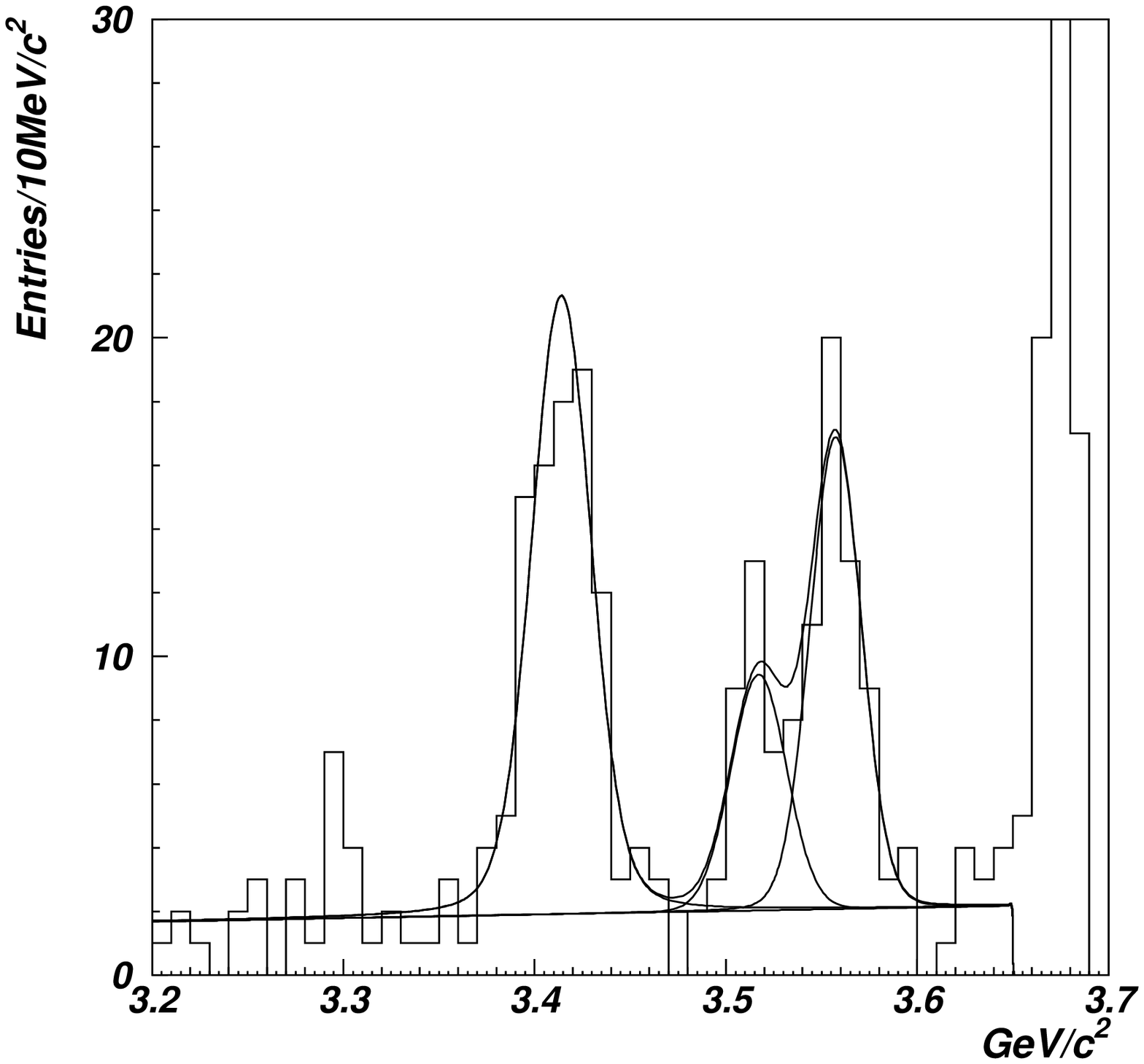,width=5.5cm}}
\caption{The $\pppr$ invariant mass distribution.
The smooth curve is the result of a fit described in the text.}
\label{mallpppr}
\end{figure}

The $\pppr$ invariant mass distribution for the selected events 
is shown in Fig.\ \ref{mallpppr}.   Here clear signals
for all three $\chicJ$ states are apparent. 
We fit the mass spectrum using the same
method described in the previous section; the results 
are listed in Table\ \ref{tab-pppr-res}
and shown as the smooth curve in Fig.\ \ref{mallpppr}.

\subsubsection{$\gamma K^+K^-K^+K^-$}

\begin{figure}[htbp]
\centerline{
\psfig{file=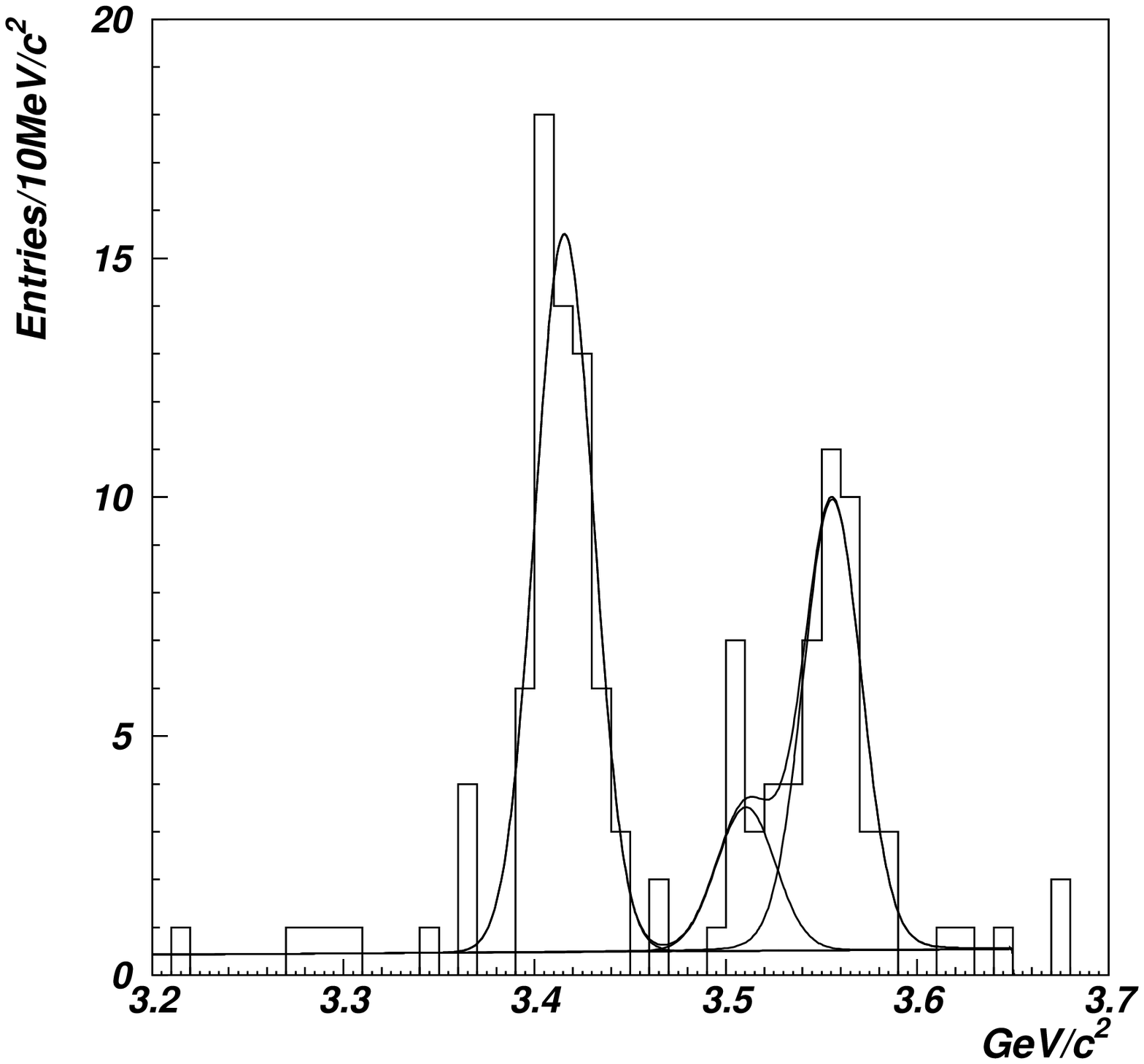,width=5.5cm}}
\caption{The $\kkkk$ invariant mass distribution.
The smooth curve is the result of a fit described in the text.} 
\label{mallkkkk}
\end{figure}

For the case where all the tracks are kaons, 
the contamination from
$\psipp$ is not an important background, and 
the requirement on total momentum of the lowest momentum $\pip$ and $\pim$
tracks, which is aimed at removing
these events, is not used.
The $\kkkk$ invariant mass distribution is shown in
Fig.\ \ref{mallkkkk}.

\begin{figure}[htbp]
\centerline{\hbox{
\psfig{file=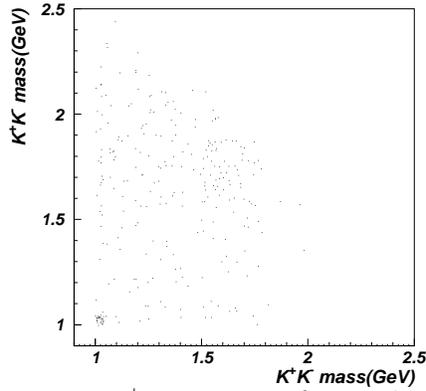,width=5.5cm}}}
\caption{A scatterplot of $\kk$ vs $\kk$ masses from
selected $\gkkkk$ events (two entries per event).}
\label{mphi-mphi}
\end{figure}

\begin{figure}[htbp]
\centerline{
\psfig{file=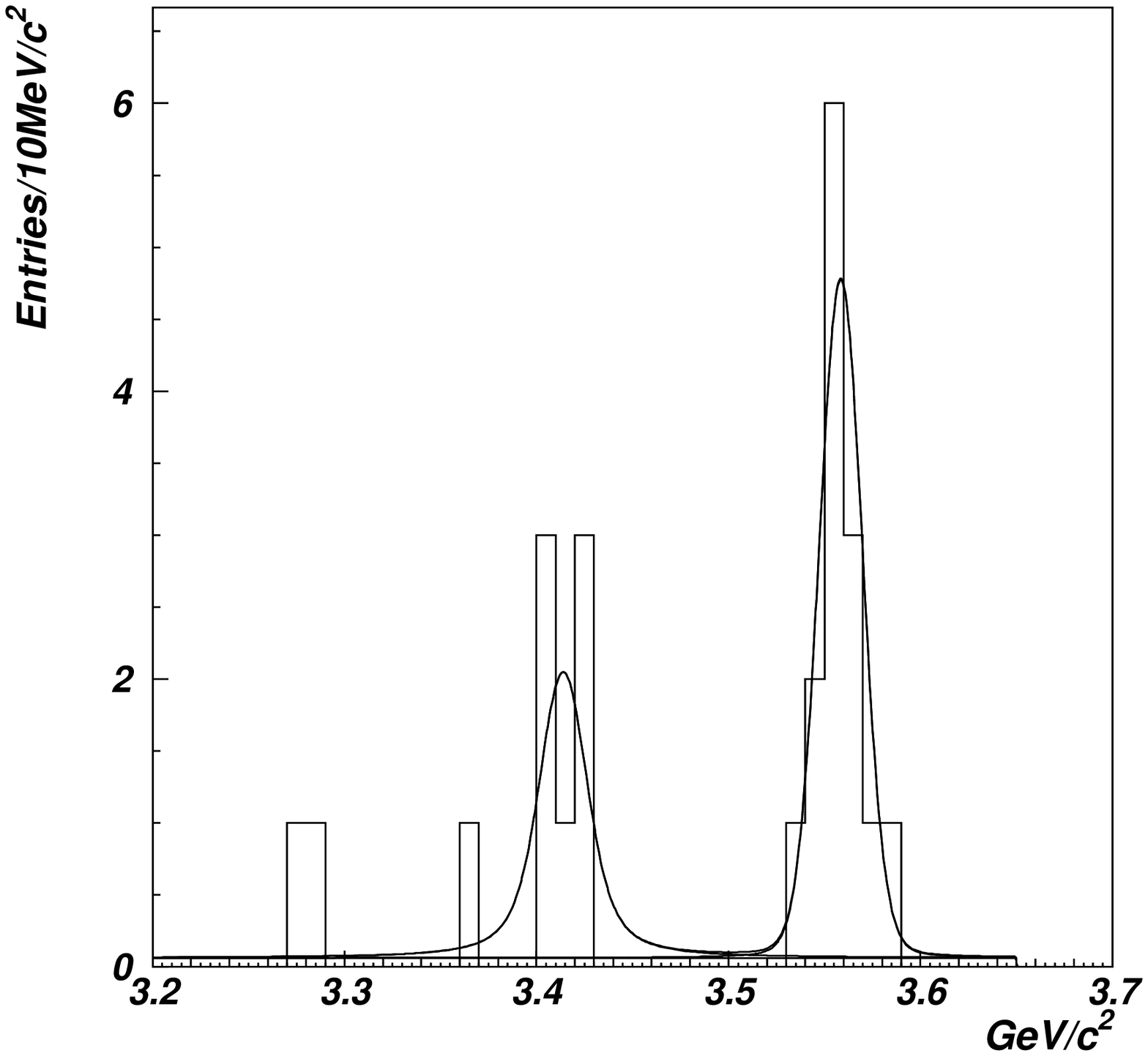,width=5.5cm}}
\caption{The $\phiphi$ invariant mass distribution.
The smooth curve is the result of a fit described in the text.} 
\label{mallphiphi}
\end{figure}

Figure\ \ref{mphi-mphi} shows a scatterplot of
$\kk$ vs $\kk$  invariant masses for the events with 
$\kkkk$ mass between 3.2 and 3.6~GeV.  The concentration  of
events in the lower left-hand corner of the plot indicates the
presence of $\phiphi$ final states.  A fit to the $\kk$ mass
distribution with a Gaussian function gives a peak mass
of $1021.9\pm 0.8$~MeV and a width 
$\sigma = 5.3 \pm 0.6$~MeV, consistent with MC expectations. 
Events where the mass of two $\kk$ combinations  
are in the range $0.99 < M_{\kk} < 1.05$~GeV
are identified as $\gamma \phi\phi$ candidates.
The $\phiphi$ mass distribution for these events is shown in 
Fig.~\ref{mallphiphi}, where there are
clear signals for the $\chi_{c0}$ and $\chi_{c2}$.

The $\kkkk$ mass and $\phiphi$ mass plots are fitted with 
three Breit-Wigner resonances and two Breit-Wigner resonances,
respectively, as described previously. The results of the
fit are listed in Table\ \ref{tab-kkkk-res} and are shown as 
smooth curves in 
Figs.\ \ref{mallkkkk} and \ref{mallphiphi}.

Because of the large fraction of $\phi\phi$
intermediate events observed in 
the $\kkkk$ mode and the significant difference
between the detection efficiency 
for phase-space events and those coming from $\phi\phi$ decays, 
the detection efficiency for the $\chi_{c0}$
and  $\chi_{c2}\rightarrow\kkkk$ channels 
is a weighted average of the phase space and 
$\phi\phi$ efficiency.
The detection efficiencies and
mass resolutions are listed in Table\ \ref{tab-kkkk-res}.
 
\subsubsection{$\gamma K^0_s K^+ \pi^- + c.c.$}

The $\chicJ\rightarrow  \kskp$ decay channels 
have serious potential backgrounds from 
$\gpppp$ (including $\gksks$) and $\gppkk$ 
final states.  To eliminate these backgrounds,
we exploit the feature  that there is one and only one 
$K^0_s$ with a secondary vertex in real $\kskp$ events.

In each event, we determine $NKSHORT$, the number of
two charged track combinations with net charge zero and effective
mass within $\pm 200$~MeV of $M_{K^{0}}$, when 
the tracks are assigned a pion mass. 
The combination with mass closest
to $M_{K^0}$ is considered to be a $K_s^0$ candidate.
The $K_s^0$ vertex is defined as the point
of closest approach of these two tracks; 
the primary vertex is defined as the
point of closest approach of the other two charged tracks
in the event.   Two parameters are used to
identify the $K_s^0$: the distance between primary vertex
and secondary vertex in the $xy$ plane, $L_{xy}$, and
the cosine of the angle between the $K_s^0$ momentum vector and
its vertex direction $CSKS$, which is expected to be very near unity
for a real $K_s^0$ event.

\begin{figure}[htbp]
\centerline{
\psfig{file=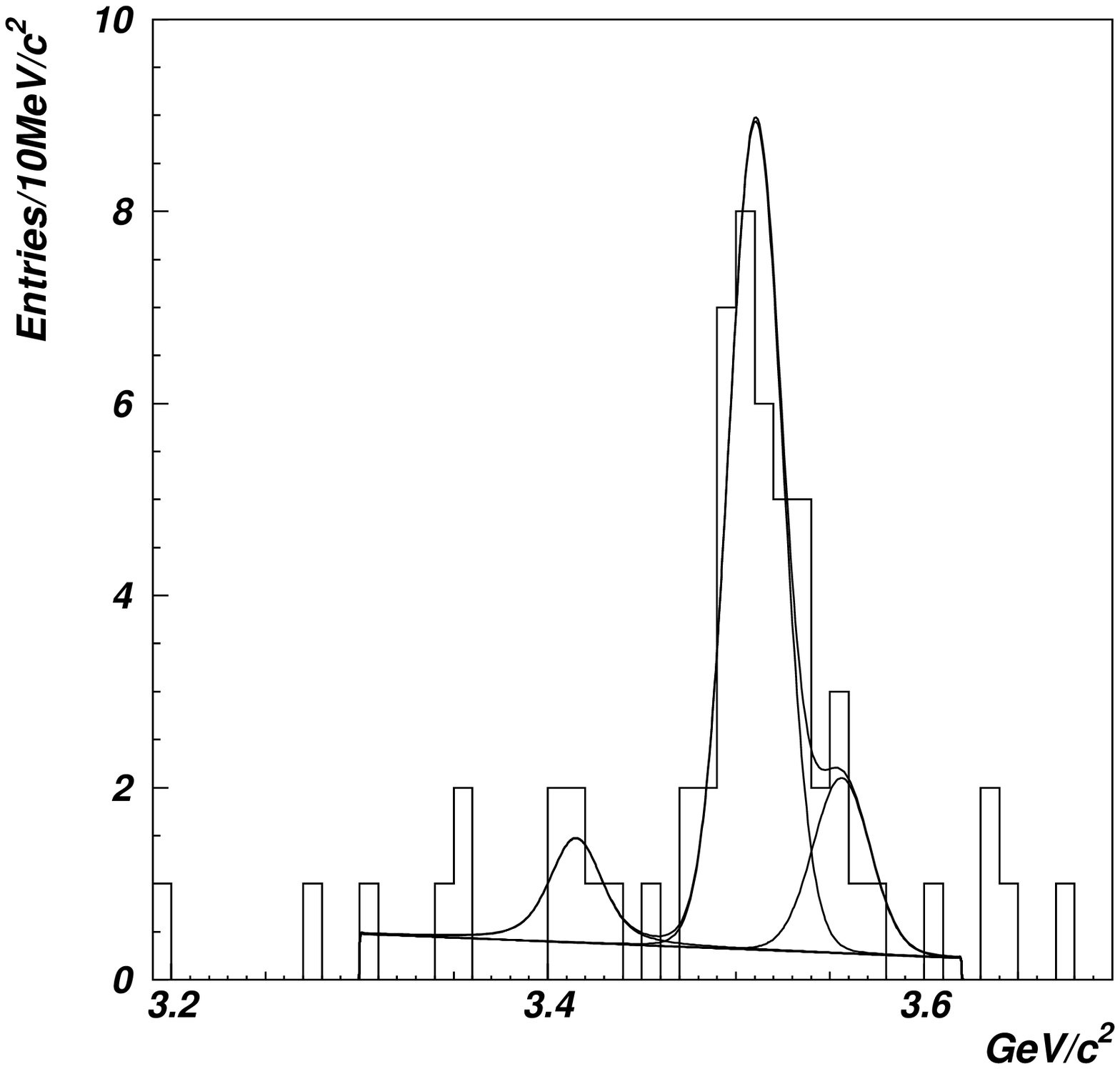,width=5.5cm}}
\caption{The $\kskp$ invariant mass distribution.
The smooth curve is the result of a fit described in the text.}
\label{mallkskp}
\end{figure}

Candidate $\gkskp$ 
events are selected by requiring the mass of the $K_s^0$ candidate 
determined from the track four-vectors returned by the
4C-fit to be within $\pm 2\sigma~({\rm i.e.}\pm 28\, \hbox{\rm MeV})$ 
of the nominal $K^0$ mass, 
$NKSHORT=1$, $L_{xy}> 5$~mm, and $CSKS> 0.98$. 
In the invariant mass distribution of the selected events, shown in
Fig.\ \ref{mallkskp}, only a $\chico$ signal is prominent.
The MC simulation indicates that the
numbers of events in the  
the $\chicz$ and $\chict$ mass region are consistent
with residual  backgrounds from
$\gpppp$, $\gksks$ and $\gppkk$ final states.
We set upper limits on the branching
fractions of $\chicz$ and $\chict$.

The $\kskp$ invariant mass distribution between 3.20 and 3.65~GeV
are fitted with the procedure described above.
The mass resolutions are fixed at their MC-determined values,
the width of the $\chi_{c0}$ is fixed at
the recent
BES value of 14.3~MeV~\cite{width} and 
those of the $\chi_{c1}$ and $\chi_{c2}$ at 
their PDG values~\cite{pdg}. 
The mass of the three $\chi_c$
states are also fixed at their PDG~\cite{pdg} values.
The fit results are listed in Table\ \ref{tab-kskp-res}
and are shown as a smooth curve in Fig.\ \ref{mallkskp}.

\subsubsection{$\gamma \tpp$}

After the selections based on the 
kinematic fit and particle ID, the main background
to the $\chicJ\rightarrow\tpp$ decays comes from
the decay chain $\pspto \psipp, \psito \gpppp$. 
The requirement
on the total momentum of the lowest momentum  $\pip$ and $\pim$ tracks
removes one third of the MC-simulated 
events while rejecting almost all the
$\psipp$ background.

\begin{figure}[htbp]
\centerline{
\psfig{file=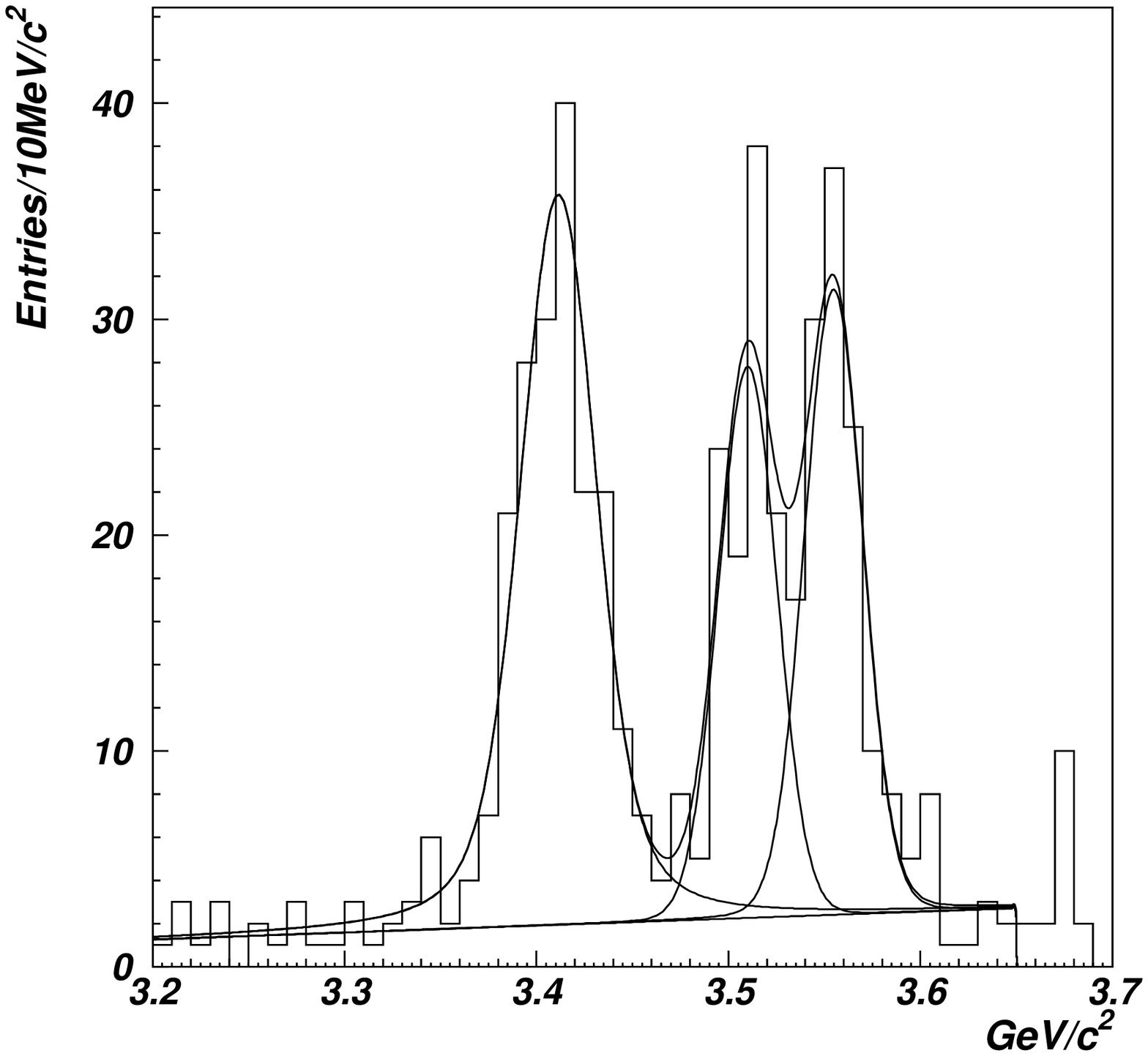,width=5.5cm}}
\caption{The $\tpp$ invariant mass distribution.
The smooth curve is the result of a fit described in the text.}
\label{malltpp}
\end{figure}

The $\tpp$ invariant mass distribution for the selected events
is shown in Fig.\ \ref{malltpp}, where prominent signals
for all three $\chicJ$ states can
be seen.   The smooth curve in the figure
is the result of the fitting procedure described above.
The results of the fit and the MC-determined
efficiencies and resolutions are listed in 
Table\ \ref{tab-tpp-res}.

\section{Branching fraction determination}

We determine branching fractions from the relation 
\[
{\cal B}(\chicJ\rightarrow X) = 
\frac{n^{obs}/\eff(\chicJ\rightarrow X)}
{N_{\psp} {\cal B}(\psp\rightarrow\gamma\chicJ)},
\]
where the values for ${\cal B}(\psp\rightarrow\gamma\chicJ)$ 
are taken from the PDG tables~\cite{pdg}.
For the $\ksks$ [$\phiphi$] channel, a factor
of ${\cal B}(K^0_s\rightarrow\pi^+\pi^-)^2$ 
[${\cal B}(\phi\rightarrow K^+K^-)^2$]
is included in the denominator.

\subsection{Systematic errors}

Systematic errors common to all modes include the
uncertainties in the total number of $\psi(2S)$
events (8.2\%) and the
$\psi(2S) \rightarrow \gamma \chi_{cJ}$ branching fractions
(8.6\%, 9.2\% and 10.3\% for $\chicz$, $\chico$ and $\chict$, 
respectively). 
Other sources of systematic errors were considered.
The variation of our results for different choices of the selection criteria
range from 10\% for high statistics channels to
25\% for those with low statistics.  The
systematic errors due to the statistical precision of the
MC event samples range from 2\% to 5\% depending on
the detection efficiencies of the channels.  Changes 
in the detection efficiency when the phase space
event generator is replaced by one  using
possible intermediate resonant states indicate 
that the systematic error on the efficiency
due to the unknown dynamics of the decay processes is 15\%.
The variation of the 
numbers of observed events due to shifts of the mass
resolutions and the total widths of the $\chi_{cJ}$ states is 
7\%;  that coming from changes in the shape used for the background
function is less than 5\%.  
The total systematic error is taken as the quadrature sum
of the individual errors and 
ranges from 25\% to 35\%, depending on the channel.

\subsection{Branching fraction results}

The branching fraction 
results are listed in Table\ \ref{chic-result}, where
all BES results for
$\chicJ$ branching fractions are given,  
including those for the two-charged track
modes reported in Ref.~\cite{width}.  In each case,
the first error listed is statistical and the second is
systematic.  For comparison, we also provide 
the previous world averages for those channels 
when they exist~\cite{pdg}.

Our branching fractions for $\chicoto \kskp$, $\chiczto \ksks$,
$\chictto \ksks$, $\chiczto \phiphi$, $\chictto \phiphi$ and
$\chicJ \ra \kkkk$ (J=0,1,2) are the first reported measurements for these
decays. The results for $\chicz$ and $\chictto \ksks$
are in agreement with the isospin prediction of the $\chicJ$
decays compared with the corresponding $\kk$ branching ratios.

For the other decay modes, signals with large statistics 
are observed and the corresponding branching fractions are
determined with precisions that are significantly
better than those of existing measurements.  Note that 
our results are consistently lower than the previous
measurements, sometimes by as much as a factor of two or more.  
We can find no obvious explanation for these discrepancies.

\section{Determination of $M_{\chi_{c0}}$ and $M_{\eta_c}$}

We determine $M_{\chi_{c0}}$ by fitting the
combined invariant mass distribution of
all of the channels discussed above to three resolution-broadened
Breit-Wigner functions with the resolution fixed at the
value of 13.8~MeV, which is determined from fits to the $\chi_{c1}$
and $\chi_{c2}$, and the total widths
of the $\chi_{c1}$ and $\chi_{c2}$ fixed at the PDG
values~\cite{pdg}. The masses of all three $\chicJ$ states
and the total width of the $\chi_{c0}$ are left as free
parameters.  The results of the fit for $M_{\chi_{c1}}$ 
($3509.4\pm 0.9$~MeV) and $M_{\chi_{c2}}$ ($3556.4\pm 0.7$~MeV)
agree with the PDG values within errors.
The fit value for $M_{\chi_{c0}}$ is $3414.1\pm0.6$~MeV,
where the error is statistical.   
The fit gives a total width for the $\chi_{c0}$ that is
in good agreement
with the recently reported BES result~\cite{width}.

\begin{figure}[h]
\centerline{
\psfig{file=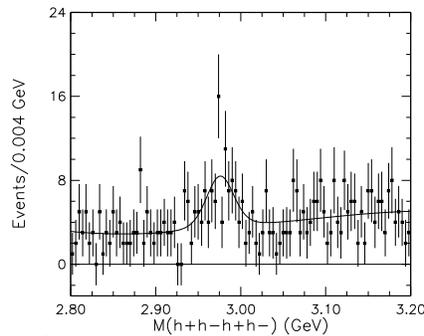,width=5.5cm}}
\caption{
The four charged track invariant mass distribution for selected events
in the $\eta_c$ mass region. The superimposed curve is the result of 
the fit described in the text.}
\label{etac_mass}
\end{figure}

Figure~\ref{etac_mass} shows the combined invariant mass
distribution for the $\pi^+\pi^-\pi^+\pi^-$,
$\pi^+\pi^-K^+K^-$, $K^+K^-K^+K^-$, and $\kskp$ channels
in the region of the $\eta_c$, where an $\eta_c$ signal is evident.
Superimposed on the plot is a fit to the spectrum using a resolution-smeared
Breit-Wigner line shape with a mass that is allowed
to vary, a total width fixed at the
PDG value of $\Gamma_{\eta_c}= 13.2$~MeV~\cite{pdg}, and a fourth-order
polynomial background function.  The
fit gives a total of $63.5\pm 14.4$ events
in the peak and has a $\chi^2/dof = 97.4/92$, which corresponds
to a confidence level of 27.9\%. The mass value from the 
fit is $M_{\chi_{\eta_c}}=2975.8 \pm 3.9$~MeV, where the error
is statistical. 
(A fit with only the background
function and no $\eta_c$ has a confidence level of 0.8\%.)

The systematic error on the mass determination includes 
a possible uncertainty in the
overall mass scale ($\pm 0.8$~MeV), which is determined
from the rms average of the differences between 
the fitted values for 
$M_{\chi_{c1}}$ and $M_{\chi_{c2}}$ and their PDG values.
The systematic errors associated with uncertainties is the
particle's total widths and the experimental resolutions 
($\pm 0.95$~MeV for $M_{\eta_c}$
and less than $\pm 0.2$~MeV for $M_{\chi_{c0}}$) are added in quadrature. 
The resulting  masses and errors are:
\[ M_{\chi_{c0}}=3414.1 \pm 0.6(\hbox{stat}) 
\pm 0.8 (\hbox{sys})~{\hbox{\ MeV}}, \]
and
\[ M_{\eta_c}=2975.8 \pm 3.9(\hbox{stat}) 
\pm 1.2 (\hbox{sys})~{\hbox{\ MeV}}. \]
The precision of our $M_{\chi_{c0}}$ measurement
represents a substantial 
improvement on the existing PDG value of $3417.3\pm2.8$~MeV~\cite{pdg}.
Our result for $M_{\eta_c}$
agrees with the DM2 group's value of $2974.4\pm 1.9$~MeV~\cite{dm2}
and is 2.4 standard deviations below the E760 group's
result of $2988.3^{+3.3}_{-3.1}$~MeV~\cite{E760}.

\section{Summary}

Events of the type $\psp\rightarrow\gamma\chicJ$ in a $3.79\times 10^6$
$\psp$ event sample are used to determine branching fractions for
$\chicJ$ decays to four and six charged particle final states.
Our results for $\kskp$, $\ksks$,
$\phiphi$, and $\kkkk$ are the first measurements for these
decays.   The branching fractions for $\chicJ\rightarrow\pppp$,
$\ppkk$, $\pppr$, and $\tpp$ final states are measured with better 
precision and found to be consistently
lower than previous measurements. 
$M_{\chi_{c0}}$ and $M_{\eta_c}$ were determined 
using the same data sample.  

\acknowledgments

We thank the staffs of the BEPC Accelerator
and the Computing Center at the Institute of High Energy Physics,
Beijing, for their outstanding scientific efforts. This project
was partly supported by China Postdoctoral Science Foundation.
The work of the BES Collaboration was supported in part by
the National Natural Science Foundation of China under Contract 
No. 19290400 and the Chinese Academy of Sciences under 
contract No. H-10 and E-01 (IHEP),   and by the Department of
Energy under Contract Nos. DE-FG03-92ER40701 (Caltech), 
DE-FG03-93ER40788 (Colorado State University),
DE-AC03-76SF00515 (SLAC), DE-FG03-91ER40679 (UC Irvine), 
DE-FG03-94ER40833 (U Hawaii), DE-FG03-95ER40925 (UT Dallas).

\begin{table}[htbp]
\caption{Fit results for $\chicJto \pppp, \ppkk$ and $\ksks$
decays.}
\begin{center}
\begin{tabular}{|c|c|c|c|}
\hline
Channel & $n^{obs}$ & $\eff$~(\%) & $\sigma_{res}$~(MeV) \\
\hline
$\chiczto \pppp$ & $874\pm 30$ & 16.06 & 15.1 \\
$\chicoto \pppp$ & $277\pm 19$ & 17.06 & 15.6 \\
$\chictto \pppp$ & $425\pm 21$ & 15.09 & 13.4 \\
\hline
$\chiczto \ksks$ & $49.3\pm 7.0$ & 15.16 & 10.9 \\
$\chictto \ksks$ & $11.7\pm 3.2$ & 13.92 &  9.4 \\
\hline
$\chiczto \ppkk$ & $587\pm 27$ & 11.32 & 14.4 \\
$\chicoto \ppkk$ & $192\pm 16$ & 12.91 & 15.3 \\
$\chictto \ppkk$ & $267\pm 18$ & 11.42 & 15.1 \\
\hline
\end{tabular}
\end{center}
\label{tab-ppkk-res}
\end{table}

\begin{table}[htbp]
\caption{Fit results for $\chicJto \pppr$ decays.}
\begin{center}
\begin{tabular}{|c|c|c|c|}
\hline
Channel & $n^{obs}$ & $\eff$~(\%) & $\sigma_{res}$~(MeV) \\
\hline
$\chicz$ & $81  \pm 10 $ & 14.62 & 13.9 \\
$\chico$ & $27.1\pm 6.9$ & 16.72 & 14.3 \\
$\chict$ & $50.9\pm 8.1$ & 13.98 & 13.0 \\
\hline
\end{tabular}
\end{center}
\label{tab-pppr-res}
\end{table}

\begin{table}[htbp]
\caption{Fit results for $\chicJto \kkkk$ and $\phiphi$ decays.}
\begin{center}
\begin{tabular}{|c|c|c|c|}
\hline
Channel & $n^{obs}$ & $\eff$~(\%)~(PS/$\phiphi$) & $\sigma_{res}$~(MeV) \\
\hline
$\chiczto \kkkk$   & $57.8\pm 6.9$ & 7.38/10.06 & 15.4 \\
$\chicoto \kkkk$   & $11.7\pm 4.2$ & 8.52/no    & 15.2 \\
$\chictto \kkkk$   & $36.6\pm 5.9$ & 7.64/9.76  & 14.7 \\
\hline
$\chiczto \phiphi$ & $ 7.6\pm 2.8$ & 9.78       &  8.9 \\
$\chictto \phiphi$ & $13.6\pm 3.7$ & 9.54       & 10.8 \\
\hline
\end{tabular}
\end{center}
\label{tab-kkkk-res}
\end{table}

\begin{table}[htbp]
\caption{Fit results for $\chicJto \kskp$ decays. The upper limits are 
90\% confidence level values.}
\begin{center}
\begin{tabular}{|c|c|c|c|}
\hline
Channel & $n^{obs}$ & $\eff$~(\%) & $\sigma_{res}$~(MeV) \\
\hline
$\chicz$ & $<8.5$        &  4.94 & 10.3 \\
$\chico$ & $31.4\pm 5.6$ &  5.64 & 14.2 \\
$\chict$ & $<10.6$       &  4.93 & 14.7 \\
\hline
\end{tabular}
\end{center}
\label{tab-kskp-res}
\end{table}
	
\begin{table}[htbp]
\caption{Fit results for $\chicJto \tpp$ decays.}
\begin{center}
\begin{tabular}{|c|c|c|c|}
\hline
Channel & $n^{obs}$ & $\eff$~(\%) & $\sigma_{res}$~(MeV) \\
\hline
$\chicz$ & $191\pm 16$ & 4.62 & 15.8 \\
$\chico$ & $ 98\pm 12$ & 5.20 & 15.0 \\
$\chict$ & $112\pm 12$ & 4.23 & 14.7 \\
\hline
\end{tabular}
\end{center}
\label{tab-tpp-res}
\end{table}
	
\begin{table}[htbp]
\caption{The $\chicJ$ hadronic decay branching fractions,
determined using
$\BR(\pspto \gamma \chi_{c0}) = (9.3 \pm 0.8)\%$,
$\BR(\pspto \gamma \chi_{c1}) = (8.7 \pm 0.8)\%$ and
$\BR(\pspto \gamma \chi_{c2}) = (7.8 \pm 0.8)\%$.
}
\begin{center}
\begin{tabular}{|c|c|c||c|}
\hline\hline
Channel	 & $n^{obs}$  & Branching Ratio  & World Average\cite{pdg}  \\\hline
$\chiczto \pp$ & $720\pm 32$ & $(4.68 \pm 0.26 \pm 0.65)\times 10^{-3}$
                                  & $(7.5\pm 2.1)\times 10^{-3}$ \\
$\chictto \pp$ & $185\pm 16$ & $(1.49 \pm 0.14 \pm 0.22)\times 10^{-3}$
                                  & $(1.9\pm 1.0)\times 10^{-3}$ \\\hline
$\chiczto \kk$ & $774\pm 38$ & $(5.68 \pm 0.35 \pm 0.85)\times 10^{-3}$
                                  & $(7.1\pm 2.4)\times 10^{-3}$ \\
$\chictto \kk$ & $115\pm 13$ & $(0.79 \pm 0.14 \pm 0.13)\times 10^{-3}$
                                  & $(1.5\pm 1.1)\times 10^{-3}$ \\\hline
$\chiczto \ppb$ & $15.2\pm 4.1$ & $(15.9 \pm 4.3 \pm 5.3)\times 10^{-5}$
                                  & $<9.0\times 10^{-4}$          \\
$\chicoto \ppb$ & $4.2\pm 2.2$  & $(4.2 \pm 2.2 \pm 2.8)\times 10^{-5}$
                                  & $(8.6\pm 1.2)\times 10^{-5}$  \\
$\chictto \ppb$ & $4.7\pm 2.5$  & $(5.8 \pm 3.1 \pm 3.2)\times 10^{-5}$
                                  & $(10.0\pm 1.0)\times 10^{-5}$ \\\hline
$\chiczto \pppp$ & $874\pm 30$ & $(15.4\pm 0.5\pm 3.7)\times 10^{-3}$
                                  & $(3.7\pm 0.7)\times 10^{-2}$ \\
$\chicoto \pppp$ & $277\pm 19$ & $( 4.9\pm 0.4\pm 1.2)\times 10^{-3}$
                                  & $(1.6\pm 0.5)\times 10^{-2}$ \\
$\chictto \pppp$ & $425\pm 21$ & $(9.6\pm 0.5\pm 2.4)\times 10^{-3}$
                                  & $(2.2\pm 0.5)\times 10^{-2}$ \\\hline
$\chiczto \ksks$ & $49.3\pm 7.0$ & $(1.96\pm 0.28\pm 0.52)\times 10^{-3}$
                                  &  \\
$\chictto \ksks$ & $ 11.7\pm 3.2$ & $(0.61\pm 0.17\pm 0.16)\times 10^{-3}$
                                  & \\\hline
$\chiczto \ppkk$ & $587\pm 27$ & $(14.7\pm 0.7\pm 3.8)\times 10^{-3}$
                                  & $(3.0\pm 0.7)\times 10^{-2}$ \\
$\chicoto \ppkk$ & $192\pm 16$ & $(4.5\pm 0.4\pm 1.1)\times 10^{-3}$
                                  & $(9\pm 4)\times 10^{-3}$ \\
$\chictto \ppkk$ & $267\pm 18$ & $(7.9\pm 0.6\pm 2.1)\times 10^{-3}$
                                  & $(1.9\pm 0.5)\times 10^{-2}$ \\\hline
$\chiczto \pppr$ & $81\pm 11$ & $(1.57\pm 0.21\pm 0.54)\times 10^{-3}$
                                  & $(5.0\pm 2.0)\times 10^{-3}$ \\
$\chicoto \pppr$ & $27.1\pm 6.9$ & $(0.49\pm 0.13\pm 0.17)\times 10^{-3}$
                                  & $(1.4\pm 0.9)\times 10^{-3}$ \\
$\chictto \pppr$ & $50.9\pm 8.1$ & $(1.23\pm 0.20\pm 0.35)\times 10^{-3}$
                                  & $(3.3\pm 1.3)\times 10^{-3}$\\\hline
$\chiczto \kkkk$ & $57.8\pm 6.9$ & $(2.14\pm 0.26\pm 0.40)\times 10^{-3}$
                      & \\
$\chicoto \kkkk$ & $ 11.7\pm 4.2$ & $(0.42\pm 0.15\pm 0.12)\times 10^{-3}$
                      & \\
$\chictto \kkkk$ & $36.6\pm 5.9$ & $(1.48\pm 0.26\pm 0.32)\times 10^{-3}$
                      & \\\hline
$\chiczto \phiphi$& $ 7.6\pm 2.8$ & $(0.92\pm 0.34\pm 0.38)\times 10^{-3}$
                      & \\
$\chictto \phiphi$& $13.6\pm 3.7$ & $(2.00\pm 0.55\pm 0.61)\times 10^{-3}$
                      & \\\hline
$\chiczto \kskp$ & $<8.5$ & $<0.71\times 10^{-3}$
                      & \\
$\chicoto \kskp$ & $31.4\pm 5.6$ & $(2.46\pm 0.44\pm 0.65)\times 10^{-3}$
                      & \\
$\chictto \kskp$ & $<10.6$ & $<1.06\times 10^{-3}$
                      & \\\hline
$\chiczto \tpp$  & $191\pm 16$ & $(11.7\pm 1.0\pm 2.3)\times 10^{-3}$
                      & $(1.5\pm 0.5)\times 10^{-2}$\\
$\chicoto \tpp$  & $ 98\pm 12$ & $(5.8\pm 0.7\pm 1.2)\times 10^{-3}$
                      & $(2.2\pm 0.8)\times 10^{-2}$\\
$\chictto \tpp$  & $112\pm 12$ & $(9.0\pm 1.0\pm 2.0)\times 10^{-3}$
                      & $(1.2\pm 0.8)\times 10^{-2}$\\\hline\hline
\end{tabular}
\end{center}
\label{chic-result}
\end{table}


\begin{thebibliography}{99}

\bibitem[\dag]{atNU0} Deceased.

\bibitem{chic_obs} W.~Braunschweig {\em et al.} (DASP Collab.),
               Phys. Lett. {\bf B57}, 407 (1975);
               G.~J.~Feldman  {\em et al.} (Mark~I Collab.),
	       Phys. Rev. Lett. {\bf 35}, 821 (1975).

\bibitem{pdg} C.~Caso {\em et al.} (Particle Data Group),
           Eur. Phys. J. {\bf C3}, 1 (1998)
	   and references therein.
		      
\bibitem{tanenbaum} W.~Tanenbaum  {\em et al.} (Mark~I Collab.),
             Phys. Rev. {\bf D17}, 1731 (1978).

\bibitem{E760a}  T.~Armstrong {\em et al.} (E760 Collaboration), 
           Phys.  Rev. Lett. {\bf 68}, 1468 (1992).
      
\bibitem{width} J.~Z.~Bai {\em et al.} (BES Collab.),
               Phys. Rev. Lett. {\bf 81}, 3091 (1998).

\bibitem{E760} T.~A.~Armstrong {\em et al.} (E760 Collab.), 
               Phys. Rev. {\bf D52}, 4839 (1995).

\bibitem{dm2} D.~Bisello {\em et al.}, (DM2 Collab.),
              Nucl. Phys. {\bf B350}, 1 (1991).

\bibitem{mk3} R.~M.~Baltrusaitis {\em et al.} (Mark~III Collab.), 
              Phys. Rev. {\bf D33}, 629 (1986); 
              J.~E.~Gaiser {\em et al.} (Crystal Ball Collab.), 
              Phys. Rev. {\bf D34}, 711 (1986);
              C.~Bagelin {\em et al.}, Phys. Lett. {\bf B231}, 557 (1989); and
              Z.~Bai {\em et al.} (Mark~III Collab.), 
	      Phys. Rev. Lett. {\bf 65}, 1309 (1990).
	       
\bibitem{npsp} J.~Z.~Bai {\em et al.} (BES Collab.),
               Phys. Rev. {\bf D58}, 092006 (1998).
	       In the determination of the number of
	       $\psp$ events, the branching ratio
	       $\BR(\pspto \psipp ) = (32.4 \pm 2.6)\%$ 
	       (R.~M.~Barnett {\em et al.}, (Particle Data Group),
	       Phys. Rev. {\bf D54} part I (1996))
	       was used. 

\bibitem{bes} J.~Z.~Bai {\em et al.} (BES Collab.),
              Nucl. Instrum. Methods Phys. Res., 
	      Sect. {\bf A344}, 319 (1994).
	      
\bibitem{e1} G.~Karl, S.~Meshkov and J.~L.~Rosner, Phys.Rev., 
                   {\bf D13}, 1203 (1976);
             M.~Oreglia {\em et al.} (Crystal Ball Collab.), 
             Phys. Rev. {\bf D25}, 2259 (1982).

\bibitem{jpsikk} Using the $\psp\rightarrow\ppkk$ events
                 sample and the particle identification procedures
                 described in the text, we determine the branching fraction 
           $\BR(\jpsi \ra \kk) = (2.35 \pm 0.34 \pm 0.44)\times 10^{-4},$
            which is in good agreement with the world average~\cite{pdg}.

\end{thebibliography}
\end{document}